\documentclass[aps,prb,twocolumn,superscriptaddress,groupedaddress]{revtex4}
\usepackage{graphicx}
\usepackage{fancyhdr}
\usepackage{amsmath}
\usepackage{xcolor}

\begin{document}

\preprint{APS/123-QED}

\title{Effects of asymmetry in strongly coupled spin vortex pairs}

\author{E. Holmgren}
\affiliation{Nanostructure Physics, Royal Institute of Technology, Stockholm, Sweden}

\author{M. Persson}
\affiliation{Nanostructure Physics, Royal Institute of Technology, Stockholm, Sweden}

\author{V. Korenivski}
\affiliation{Nanostructure Physics, Royal Institute of Technology, Stockholm, Sweden}

\date{\today}

\begin{abstract}
Effects of magnetic asymmetry on strongly coupled spin-vortex pairs with parallel core polarization and antiparallel chirality in synthetic nanomagnets are investigated. This includes vortex-core length asymetry, biasing field asymmetry, and pinning of one of the two vortex cores. Our experimental observations as well as analytical and micromagnetic modeling show how magnetic asymmetry can be used to differentiate magneto-resistively otherwise degenerate multiple stable states of a vortex pair. These results expand the knowledge base for spin vortex arrays in nanostructures and should be useful in light of the recent proposals on coding information into multiple topological spin states, such as single and multiple vortex core/chirality states.
\end{abstract}

\maketitle

\section{\label{sec:intro}Introduction}
Spin vortices in magnetic nanostructures are widely considered as promising for applications such as magnetic memory and high-frequency oscillators. As an example, a recent proposal for a vortex-based memory\cite{Araujo2016} is based on a pinned spin vortex, which is switched between its various stable states thermo-magnetically, with the device operation highly sensitive to the specifics of the field-response by a given vortex configuration, including such effects as nanoscale Barkhausen noise.\cite{Burgess2013}  Static fields mainly interact with the in-plane circularly magnetized periphery of the vortex, the direction of which is characterized by the vortex chirality -- clockwise or counterclockwise.

Vortex-based oscillators\cite{Pribiag2007} utilize the dynamics of a spin vortex, expressed via the motion of the vortex core. The core is a small region inside the vortex with out-of-plane magnetization and is polarized``up" or ``down".\cite{Wachowiak2002} Some suggested memory implementations utilize the fact that a vortex core reversal occurs (e.g., from up to down) during its oscillation under a high-amplitude resonant excitation, which is used for writing binary information into the core's polarization state.\cite{Pigeau2010}

Coupled vortices have been shown to possess properties distinct from those of individual isolated vortices. Multi-vortex structures have been studied in several regimes of coupling, such as magnetostatic coupling in lateral arrays of magnetic particles,\cite{awad1,Galkin2006,Awad2010} magnetostatic coupling including a strong core-core interaction in vertically stacked nanoparticles,\cite{cherepov1,lebrun1,Haenze2016} and exchange-coupled vortices within the same thin ferromagnetic particle.\cite{hata1} It has been shown that the inter-vortex interactions can drastically change both static and dynamic behavior of the system. It has additionally been shown that both isolated vortices with asymmetry\cite{Compton2010} and asymmetric vortex pairs\cite{Stebliy2017} have significantly modified properties compared to their symmetric counterparts, and that magnetic asymmetry provides a rich tool-set for tuning the properties, both static and dynamic, of any vortex system.

\begin{figure}
\includegraphics[width=3.4in]{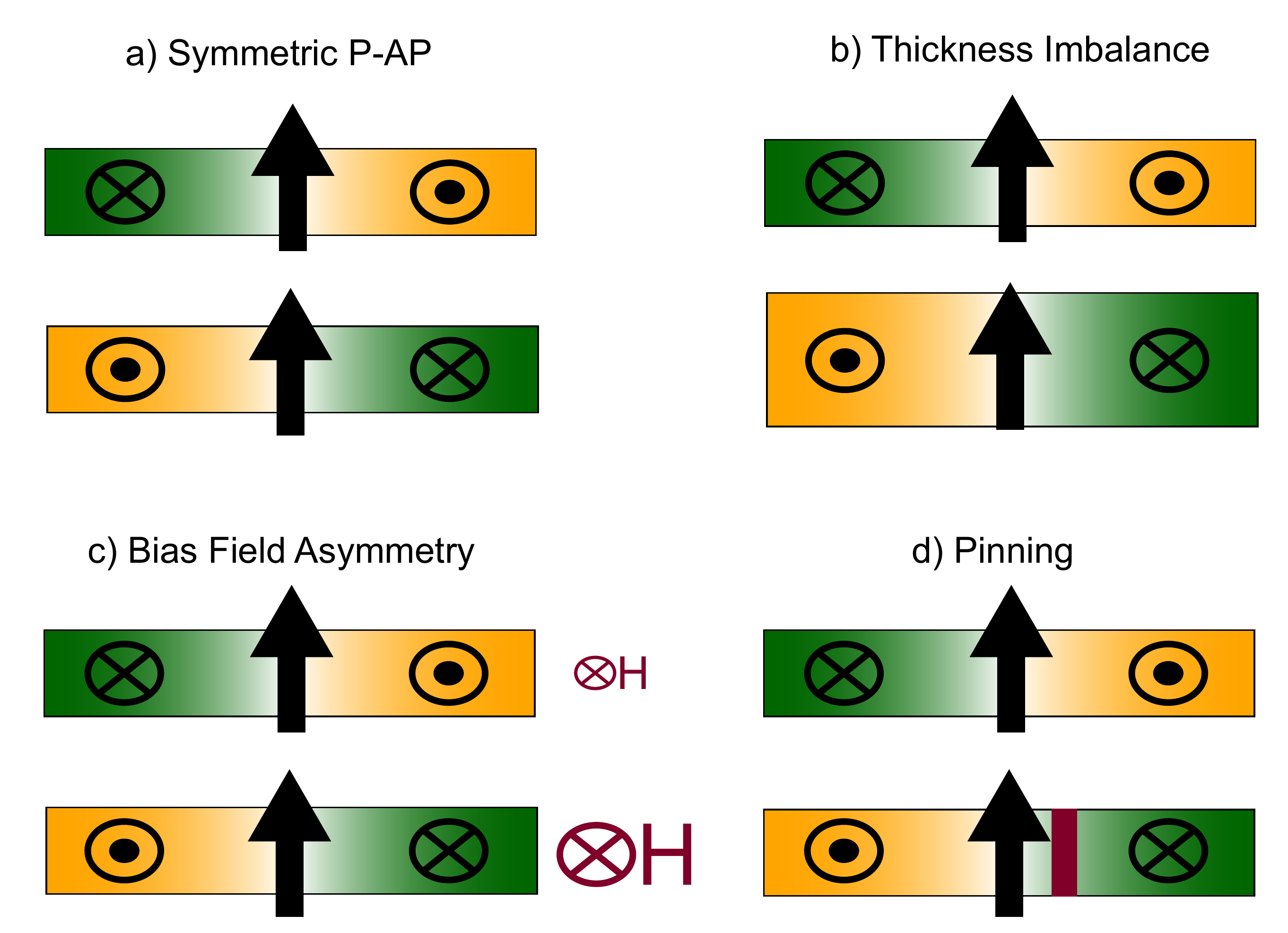}
\caption{Schematics of asymmetry that can be present in vortex pair with parallel core polarizations (P) and antiparallel chiralities (AP). Green/yellow color and in/out-of-plane periphery-spin projections represent chirality; vertical arrows - core polarization. (a) Symmetric vertical P-AP vortex pair, with two identical ferromagnetic nanoparticles in close proximity. Asymmetric spin vortex pairs with: (b) thickness imbalance, where two magnetic layers have different vortex core length, (c) bias field asymmetry, where static field $H$ acting on two vortices is different, (d) pinning of one of two vortex cores (pinning site depicted in red).}
\label{vortexstates}
\end{figure}

\begin{figure}
\includegraphics[width=0.4\textwidth]{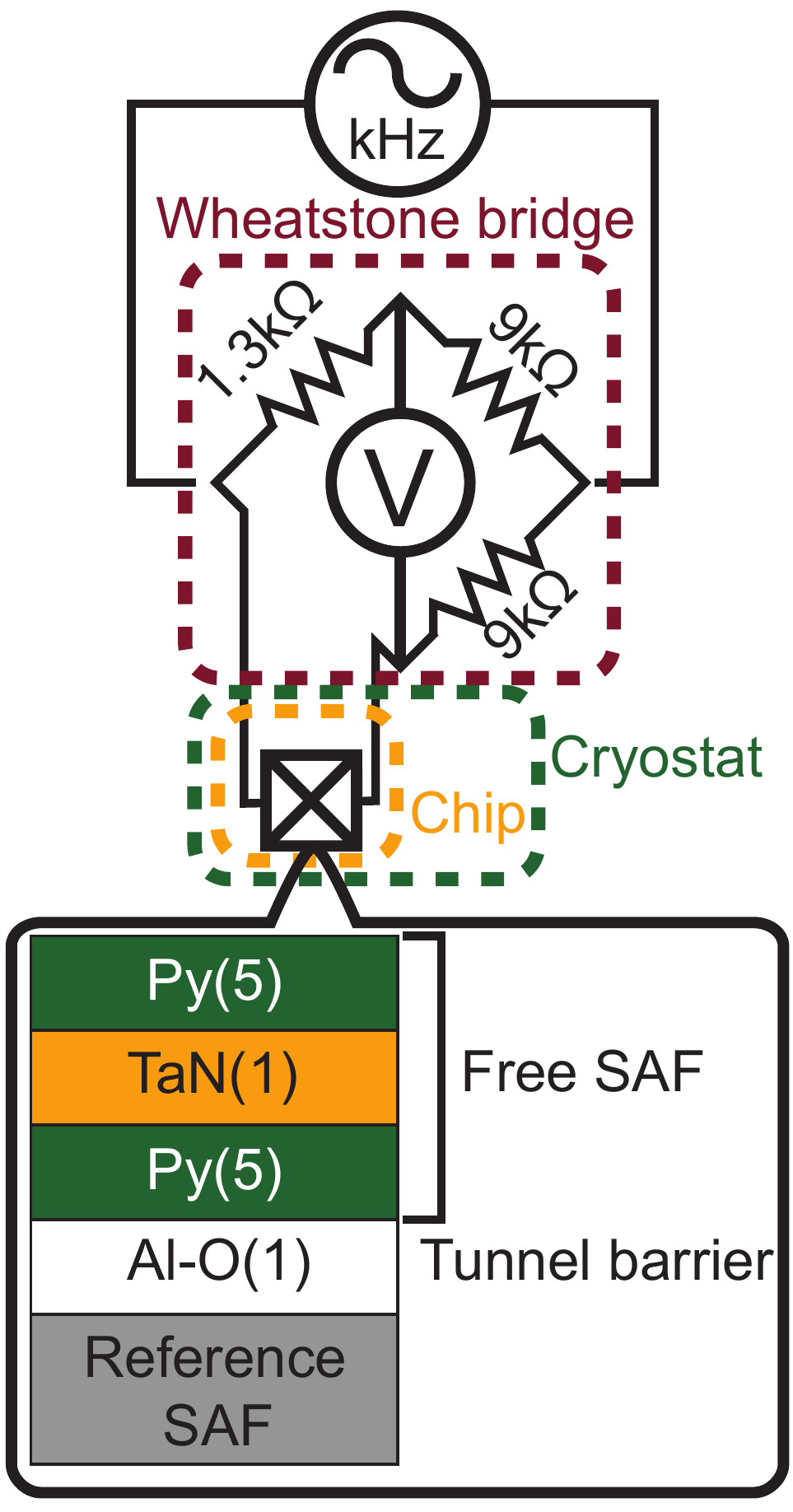}
\caption{Measurement setup and junction design. The resistance of the sample was measured with a lock-in amplifier using a Wheatstone bridge where a 1.3k~$\Omega$ resistor, being an average vortex state resistance, was used as the balance to the junction. The vortex pair is created in the two layers of the free SAF. The layers consist of permalloy and only interact magnetostatically.}
\label{setup}
\end{figure}

\begin{figure*}
\includegraphics[width=0.8\textwidth]{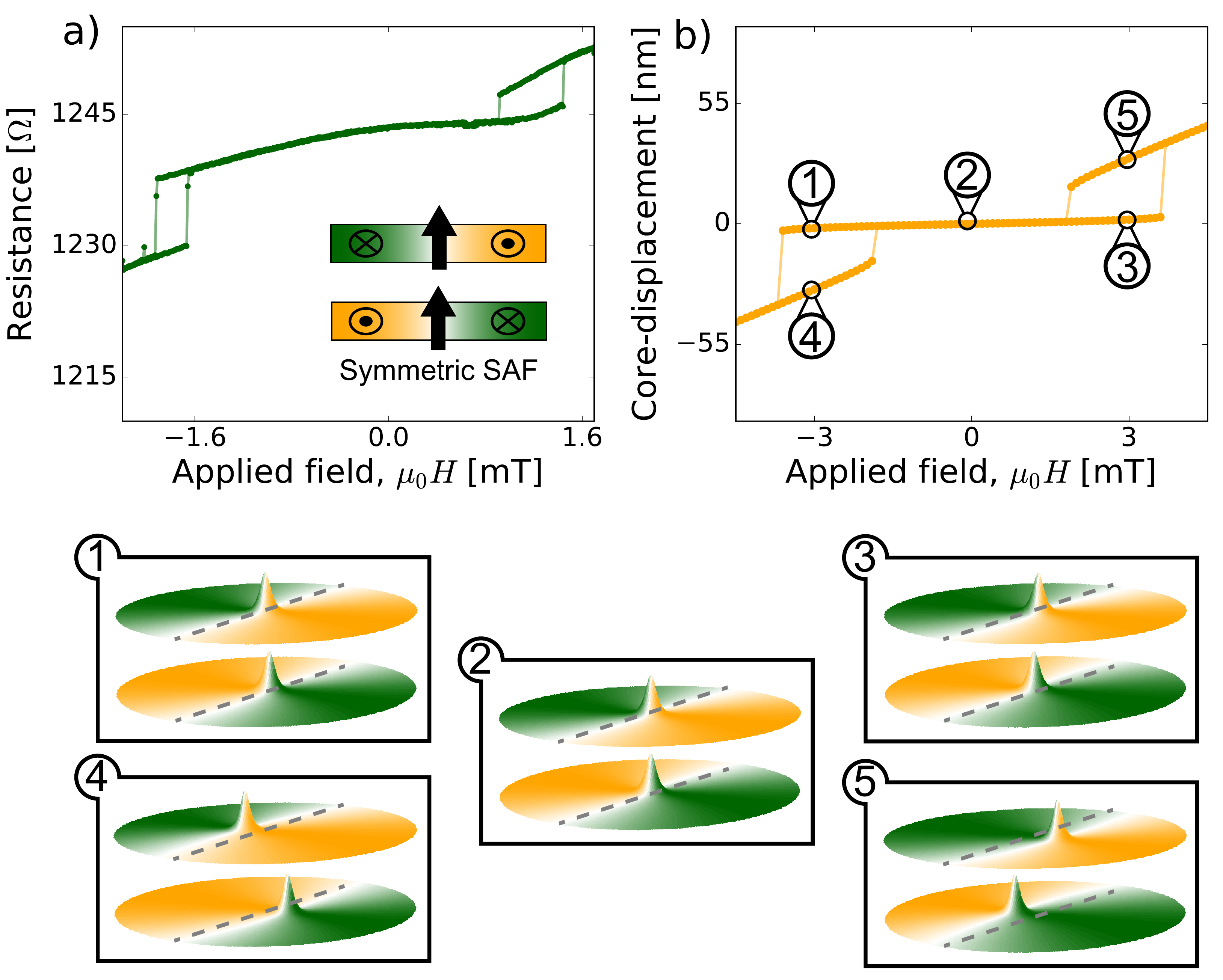}
\caption{(a) Measured core coupling/decoupling hysteresis loop of symmetric P-AP vortex pair, with inset showing geometry of vortex P-AP pair in its core-core coupled state. (b) Micromagnetic simulation for perfectly symmetric vortex pair, showing displacement of bottom core as external field is swept through hysteresis. Bottom panels, labeled 1-5, show spin distribution at different points in simulated sweep, with out-of-plane height representing z-polarizations (+1 for both vortex cores in P-AP state) and color representing magnetization component (yellow for +1 and green for -1) along easy axis (grey dashed line). For symmetric P-AP pair, positive and negative core-core decoupling fields are equal in magnitude, roughly $\pm$3.5 mT in simulations, as is off-center displacement of two cores after decoupling.}
\label{staticPAPsym}
\end{figure*}

We have previously shown that in magnetostatically-coupled vertically-stacked vortex pairs, in the limit of strong monopole-like core-core interaction, the properties of the pair are determined by the relative core polarizations and chiralities of the vortices constituting the pair.\cite{cherepov1} Of special interest is the vortex pair, where the core polarizations are parallel (P), while the chiralities are antiparallel (AP), shown schematically in Fig.~\ref{vortexstates}(a). The parallel core polarization results in a magnetostatic attraction between the two cores, which couples the vortices into a strongly bound pair-state. Due to the antiparallel chiralities the individual vortices of the pair are pulled in opposite directions by static fields leading to decoupling of the cores above some threshold field. In the coupled state, the P-AP vortex pair possesses a collective dynamic mode, not present in the isolated vortex state or the decoupled state of the pair, where the two cores in the bound pair rotate in anti-phase about the 'magnetic center' of the bi-layer particle, at a frequency an order of magnitude higher than that of the gyrational mode for isolated vortices.

In this work we show how the properties of the P-AP vortex pair are effected by magnetic asymmetry within the pair. The types of asymmetry considered are a thickness imbalance, where the two vortices have different relative weight; bias field asymmetry, where the static field acting on the two vortices are unequal; and pinning of one of the cores, which, in the coupled state of the P-AP vortex pair, effectively pins the second core and the pair as a whole.

\subsection{\label{sec:samples}Samples and micromagnetic simulations}
The samples studied in this work are elliptical nanopillars fabricated using methods described in detail elsewhere.\cite{samples1,samples2} The measurement setup and sample layout is shown in Fig.~\ref{setup}. All measurements were performed at 77 K, where the two branches of the measured hysteresis are stable for several hours. The nanopillars consist of a soft permalloy (Py, Ni$_{80}$Fe$_{20}$) synthetic antiferromagnet (SAF) separated from a reference SAF by an aluminum-oxide tunnel barrier. The long axis (or easy axis, EA) of the patterned elliptical Py particles ranged from 350 to 450 nm, with the in-plane aspect ratio of 1.2. The free SAF consists of Py(5~nm)/TaN(1~nm)/Py(5~nm), in which the direct and indirect interlayer exchange coupling are fully suppressed by the tantalum nitride spacer. The top reference layer is uniformly magnetized along the EA of the bottom free layer. As the vortex core is moved along the in-plane short axis (hard axis, HA) of the pillar the magnetization along the positive direction of the EA in the free layer increases or decreases depending on the direction of the movement, thereby changing the resistance of the pillar and allowing the position of the bottom vortex core to be read out resistively. The nanopillars are integrated on-chip with electrically isolated high-frequency waveguides.\\

The micromagnetic simulations were carried out using the MuMax3 package.\cite{Vansteenkiste2014} The standard cell size used was \{x,y,z\}=\{2.1875,2.1875,2\} nm with the total of 210$\times$175 lateral cells, and the SAF geometry was set to Py(4 nm)/TaN(2 nm)/Py(4 nm). No thermal agitation was included in the simulations. For all types, magnetic asymmetry was introduced via the bottom layer. In the case of thickness imbalance, the cell size in the z-direction was reduced to 1~nm. Pinning sites were simulated by reducing the saturation magnetization in a small cylindrical region in the bottom layer, with the center of the cylinder defining the pinning site position. This models pinning due to grain boundaries or non-magnetic inclusions, where normally the saturation magnetization is reduced.\cite{OHandley1999,Burgess2014} Unless otherwise stated, the simulation results are presented by the position of the bottom vortex core (qualitatively same behavior is found for the top vortex core). The material parameters used were the standard permalloy parameters: $M_s= 8.3\cdot 10^5$ A/m, A=$1.3\cdot 10^{11}$ J/m, $\alpha$=0.013, and no intrinsic anisotropy.

\subsection{\label{sec:pap} Properties of symmetric P-AP vortex pairs}

The two permalloy nanoparticles in the SAF can be set into their vortex states, with the total of sixteen combinations of possible and stable relative chirality and core polarization. For an ideally symmetric SAF, four non-degenerate combinations exist: parallel core polarization and anti-parallel chirality (P-AP), shown in Fig.~\ref{vortexstates}(a), AP-AP, P-P, and AP-P (last three not shown). Each combination has four degenerate configurations of individual polarizations and chiralities, physically equivalent and mutually related by basic symmetry operations.

The core-core interaction in the limit of small vertical separation between the vortices is monopole-like and the relative core polarization determines the sign of the interaction for on-axis core alignment. If the polarizations are parallel the cores strongly attract. Antiparallel cores strongly repel on-axis and weakly attract off-axis. An externally applied static magnetic field interacts predominantly with the spins in the individual vortex-periphery regions (characterized by chirality), which by volume are much larger than the vortex cores (a factor of 1000 in our case). The relative chirality of a vortex pair determines whether the vortex cores can be decoupled by the effect of an external field: an application of field moves the cores into opposite directions for anti-parallel chiralities while moving both cores as a couled pair in the same direction for parallel chiralities.\\

The P-AP vortex-pair configuration has static and dynamic properties, which are most distinct from those for the individual vortices comprising the pair.\cite{cherepov1} The potential energy of the core-core pair has four contributions:
\begin{eqnarray}
U_{tot}=\sum_{i=1,2} \left(U_{ex,i}+U_{ms,i}+U_{Z,i} \right)+ U_{c-c}
\label{Utot}
\end{eqnarray}
-- the exchange, magnetostatic, and Zeeman energy of the individual vortices in applied field, and the core-core interaction, respectively. 

For small deviations from the center of the particle, the vortex can be considered undeformed and its exchange energy component constant. The magnetostatic energy due to the particle boundary acts as a restoring force centering the individual vortex cores within the respective particle, and can for layer $i$ be written as a function of the core deviation off-center, $\mathbf{x}=(x_i,y_i)$, as\cite{Guslienko2002} 
\begin{eqnarray}
U_{ms,i}=\frac{k_x}{2}x_i^2+\frac{k_y}{2}y_i^2+\frac{k'_x}{2}x_i^4+\frac{k'_y}{2}y_i^4+2\sqrt{k'_x k'_y}x_i^2y_i^2,
\label{Ums}
\end{eqnarray}
where $k_j=20\mu_0M_s^2L_z^2/9L_j$, $k'_j=2k_j/L_j^2$ are the spring constants of the boundary terms, with $L_j$ being the length of side $j=x,~y$, and $L_z$ being the height of the core. 

The Zeeman energy can be written as
\begin{eqnarray}
U_{Z,i}=\frac{\pi}{2}\mu_0M_sL_z\zeta_i \left( H_yL_yx_i-H_xL_xy_i \right),
\end{eqnarray}
where $\zeta_i=\pm 1$ denotes the chirality. The Zeeman force shifts vortices from the center, perpendicular to the field and forces the cores apart due to the anti-parallel chirality of the vortex pair. 

The core-core interaction is a monopole-like attraction, which can be written as the sum of the pair-wise interactions of the four surface poles of the two vortex cores:\cite{cherepov1}
\begin{eqnarray}
\begin{aligned}
U_{c-c}=\mu_0 M_s^2 \Delta^3 \left[-\Phi \left( \frac{X_d}{\Delta}, \frac{D}{\Delta} \right)+\Phi \left( \frac{X_d}{\Delta}, \frac{D+L_{z,1}}{\Delta} \right) \right.	 \\ \left. +\Phi \left( \frac{X_d}{\Delta}, \frac{D+L_{z,2}}{\Delta} \right) -\Phi \left( \frac{X_d}{\Delta}, \frac{D+L_{z,1}+L_{z,2}}{\Delta} \right) \right].
\end{aligned}
\label{Ucc}
\end{eqnarray}
Here $D$ is the spacer thickness, ${X_d}$ the lateral separation, and $\Delta \approx 10$~nm - the core size in Py. The dominant first term is the attraction of the two poles adjacent to the thin spacer. Function $\Phi(d,\delta)$ is a universal function describing the potential between two charged surface poles,
\begin{eqnarray}
\Phi (d,\delta)=\frac{\pi}{4}\sqrt{2}e^{-2d^2} \int_0^{\infty} \frac{r dr}{\sqrt{r^2+\delta^2/2}}e^{-r^2}I_0(2d\sqrt{2}r),
\end{eqnarray}
where $I_0(x)$ is the modified Bessel function of the first kind.

\begin{figure*}
\includegraphics[width=0.8\textwidth]{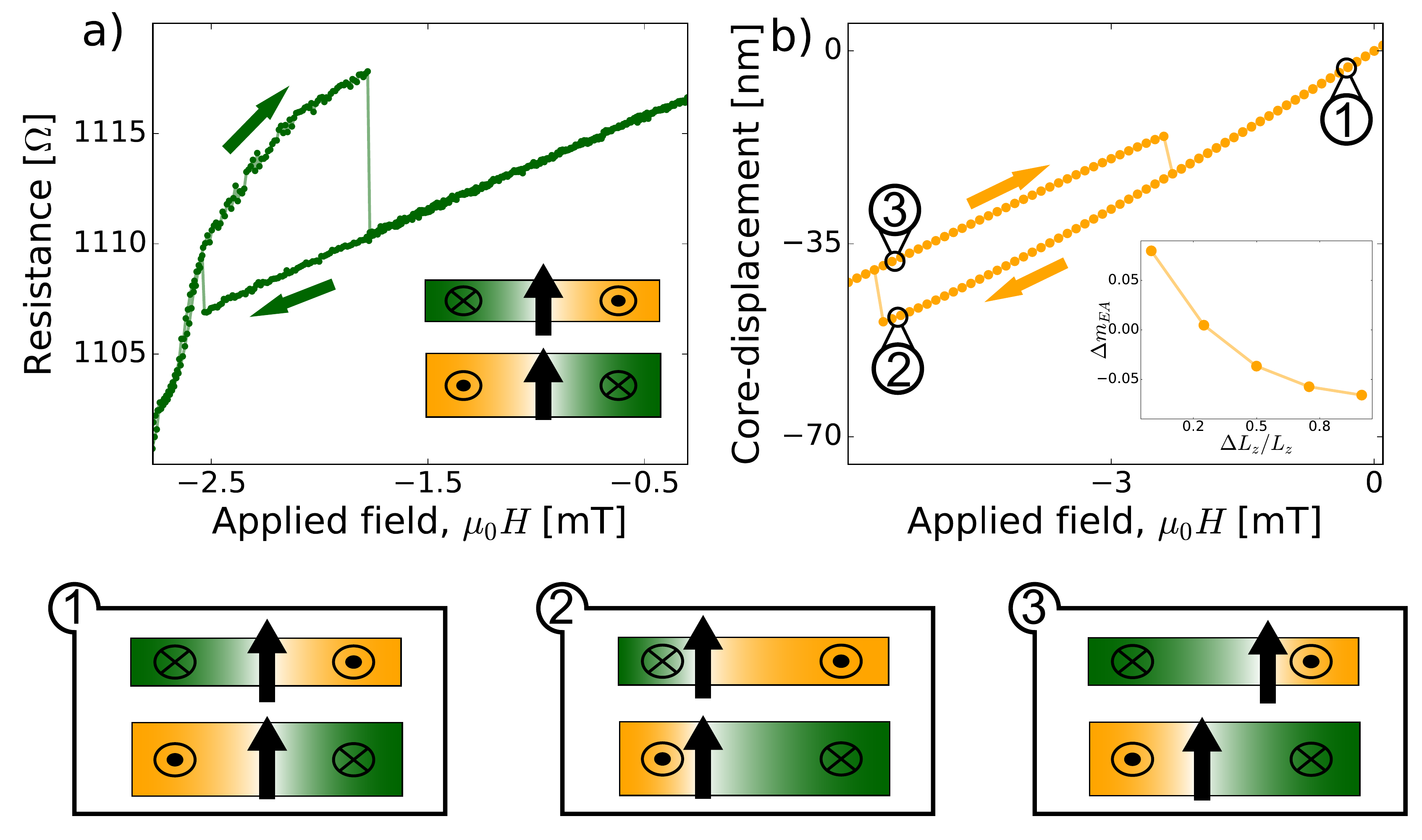}
\caption{(a) Measured core-core de/coupling hysteresis for sample with pronounced thickness imbalance. (b) Micromagnetic simulation where bottom layer is 2 nm thicker; inset shows depth of hysteresis at decoupling, $\Delta m_{EA}$, versus increasing bottom layer thickness, $\Delta L_z/L_z$. Panels 1-3 schematically show core behavior at different points during field sweep. For thickness imbalance exeeding 1 nm, as cores decouple, thicker core is displaced in opposite direction to its displacement in symmetric SAF case. Thicker core behaves more like isolated vortex core and yields smaller change in resistance upon decoupling. Thinner core has larger amplitude motion, bipolar along HA, resulting from drag-and-release by thicker core. Decoupling field is increased but is symmetric, $\pm$2.5 mT, for measured junction.}
\label{staticPAPti}
\end{figure*}

In low applied fields, the core-core attraction strongly couples the pair into a bound state, with the lateral separation between the cores much smaller than the core size. The cores remain coupled until the Zeeman energy overcomes the core-core attraction, forcing the individual vortices apart and decoupling the pair. The vortices remain decoupled until the field is reduced to a substantially smaller value, where the cores again fall within the mutual attraction potential and couple into a pair. As a result, such field sweep produces hysteresis in the core position versus the applied field. If no asymmetries are present, the negative and positive de/coupling fields as well as the off-center core displacements are equal in magnitude.

The resistance of a symmetric SAF junction in a P-AP vortex pair state as a function of applied field is shown in Fig.~\ref{staticPAPsym}(a) and compared with a micromagnetic simulation shown in Fig.~\ref{staticPAPsym}(b). Panels 1 to 3 show the spin configurations at the key points in the hysteresis loop. The discrepancy in the switching field magnitudes between the experiment and the simulations is caused by non-zero thermal agitation present on the experiment (at 77~K). The junction shows slight asymmetry expressed in the slightly different core decoupling fields, -1.8 and 1.5 mT, as well as the asymmetric in height resistance steps at core de/coupling. The source of the asymmetry is a non-ideally compensated stray field from the reference layer, estimated from the micromagnetic modeling in Fig.~\ref{staticPAPbf} to be less than 1 mT. More quantitatively, the bias-field asymmetry can be obtained from the switching fields in the uniform state of the junction.\cite{Koop2014} A separate such calibration measurement was performed and yielded a bias field asymmetry of the junction measured of approximately 0.2~mT.\\

\section{\label{sec:asymstat} Effects of Magnetic Asymmetry}
Any asymmetry within the nanoparticles modifies the magnetic properties of the P-AP vortex pair. Three types of asymmetry are analyzed in this work -- thickness imbalance, where one SAF layer is thicker than the other; bias field asymmetry, where the effective field acting on the two vortices is different (i.e., due to uneven stray fields from the reference layer); and pinning, where one core is pinned by a morphological or magnetic defect.

\subsection{Thickness imbalance}

Thickness imbalance occurs when one of the vortices resides in a thicker magnetic disk, produced either intentionally in the design of the stack or due to fabricational imperfections (e.g., often present thin "magnetically dead layers" near interfaces).\cite{Jang2010}

The direct core-core coupling becomes stronger as the thickness of one of the layers increases, caused by a reduction of the dipole-compensating contribution from the core-pole at the outer interface of the thicker layer, as described by Eq. \eqref{Ucc}. In an applied field, the higher Zeeman energy of the thicker layer makes the pair easier to move off center in the direction of the core displacement in the thicker magnetic layer, so the pair has a larger offset prior to core-decoupling. The pair is not offset at zero applied field in this configuration, however. Above a certain threshold value in the thickness imbalance, roughly 1~nm as estimated from the simulation results shown in the inset to Fig.~\ref{staticPAPti}(b), the thicker core can in fact snap back fractionally toward the particle center upon the core-decoupling.

Fig.~\ref{staticPAPti} shows the measured and simulated core-core hysteresis for a thickness-imbalanced P-AP vortex pair, for the case where the thickness increase is greater than the above mentioned threshold. The change in resistance upon decoupling is roughly $\Delta R \approx 3~\Omega$. Translating this into magnetization, with 1000$~\Omega$ in the low resistance state and 20\% MR, gives $\Delta m_{EA} = 0.03$. Comparing this to the simulation results shown in the inset to Fig.~\ref{staticPAPti}(b), yields a thickness imbalance of about 1 nm, which is reasonable for the samples studied (nominal individual Py layer thickness is 5~nm).

\subsection{Bias field asymmetry}

\begin{figure*}
\includegraphics[width=0.8\textwidth]{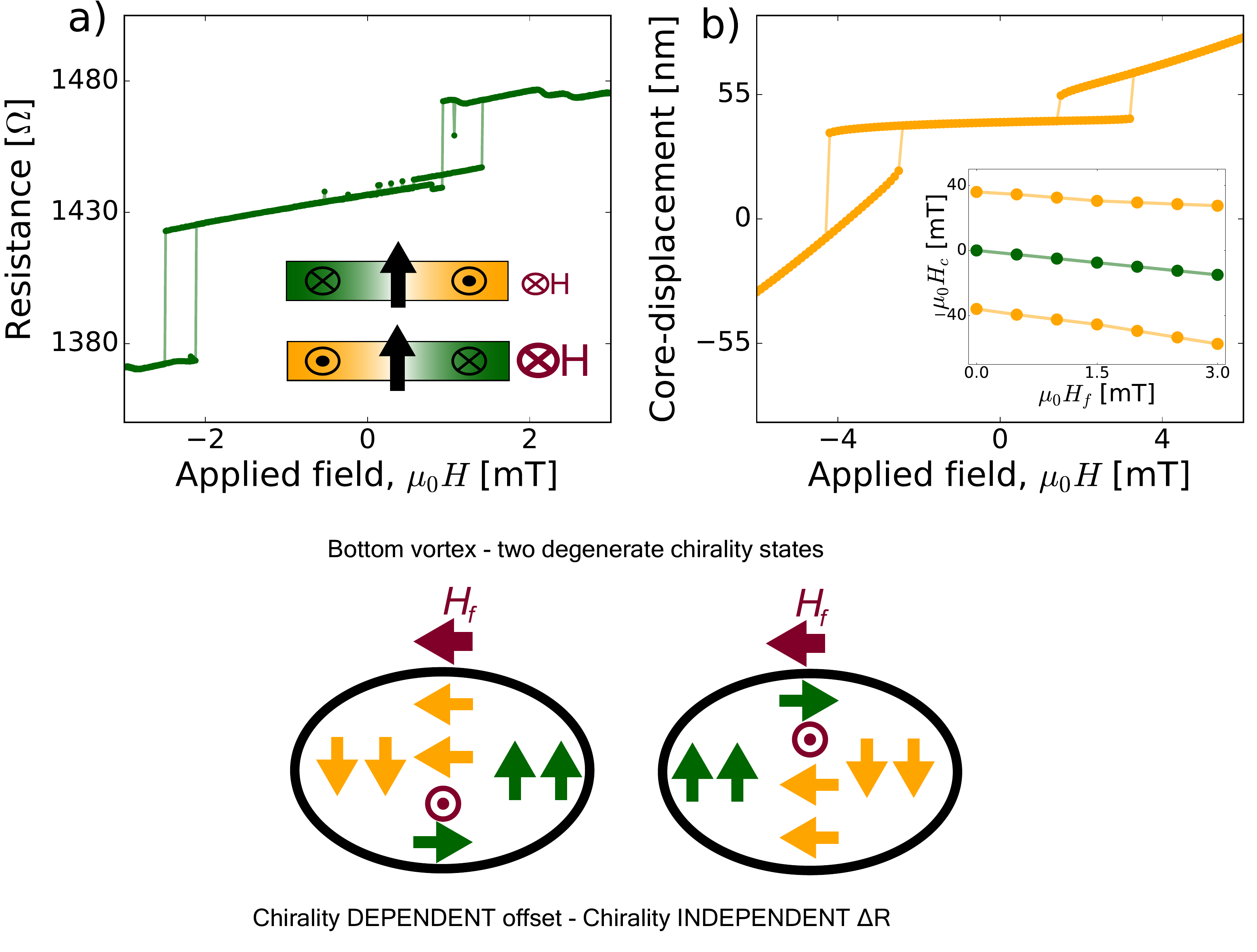}
\caption{(a) Core coupling hysteresis of sample with bias field asymmetry, with inset schematically showing relevant spin configuration. (b) Micromagnetic simulation with easy axis field higher by 1 mT in bottom Py layer versus top Py layer; inset shows positive and negative decoupling fields ($\mu_0 H_d$, yellow) as well as half difference of two decoupling fields (green) as function of magnitude of field asymmetry, $\mu_0 H_f$. Entire hysteresis is shifted in direction of offset field and magnitude of shift is fraction of total field asymmetry. Individual cores are offset from nanoparticle center at zero applied field. Offset is chirality-dependent, nevertheless, it does not result in multiple distinguishable P-AP states, as illustrated in bottom panel.}
\label{staticPAPbf}
\end{figure*}

Bias field asymmetry is a configuration, in which the effective static field acting on the two vortices is different. This can occur due to a non-ideally flux-closed reference SAF due, for example, to a thickness imbalance in the reference layers or N{\'e}el (orange-peel) coupling between the reference layer and the two-vortex SAF mediated by interface roughness.\cite{Tegen2001,Worledge2007}

$R$-$H$ hysteresis loop measured for the P-AP vortex-pair state of a junction with significant bias field asymmetry and the corresponding results of a micromagnetic simulation are shown in Fig.~\ref{staticPAPbf}. The de/coupling hysteresis is shifted in field by an amount proportional to the bias-field difference. The core decoupling in the simulation occurs at -4.2 and 3.2 mT, resulting in a loop offset of 0.5 mT for the 1 mT biasing-field asymmetry used. The loop offset for the measured junction is approximately 0.4 mT, with the core decoupling taking place at -2.4 and +1.6 mT. The comparison allows us to infer an asymmetric bias field of approximately 1 mT (even though the externally applied field is uniform). 

The hysteretic changes in the vortex magnetization, expressed in the measured junction resistance, are asymmetric in the applied field strength as well as the magnitude as the cores decouple (recouple). This is due to the bias-field asymmetry, offsetting the core-core pair off the particle center at zero applied field. It is important to note that, even though vortex pairs with opposite AP-chiralities (e.g., having the bottom vortex of clockwise or counterclockwise chirality in AP1 and AP2 states) would be offset in opposite directions, the in-plane magnetization along the bias field direction would increase in both cases, as illustrated by the schematic in Fig.~\ref{staticPAPbf}. As a result, the change in the junction resistance is the same regardless of the specific chirality of the bottom vortex, so introducing bias-field asymmetry cannot be used to distinguish the individual, nominally degenerate configurations of the P-AP state [four degenerate P-AP states in a symmetric SAF: P1(P2)-AP1(AP2)]. To summarize the effects of this type of magnetic asymmetry, neither a bias field as such or an asymmetric bias field can be used to read out multiple degenerate vortex-pair memory states.

\subsection{Pinning}

\begin{figure}
\includegraphics[width=3.1in]{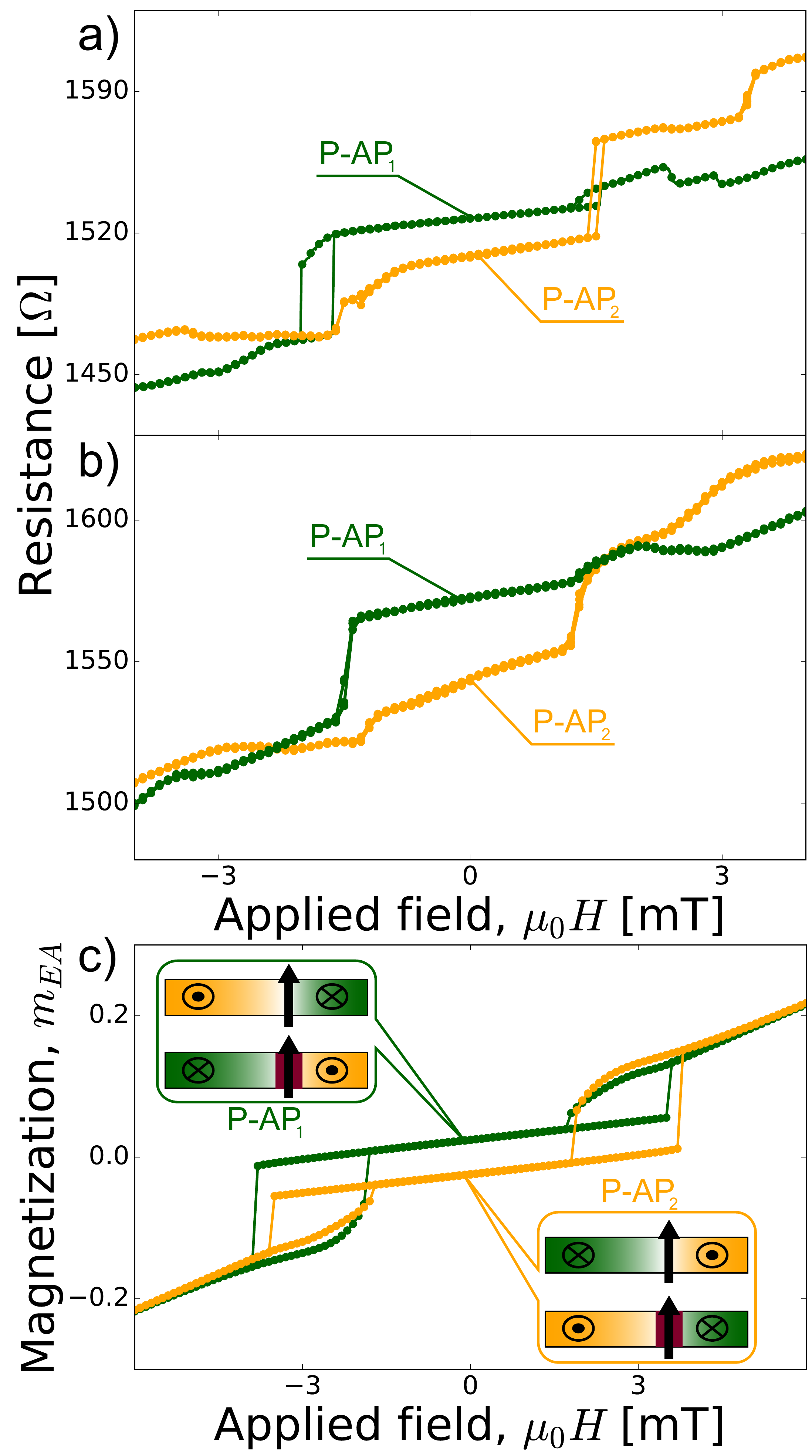}\\
\caption{Measured field sweeps at (a) 77 K and (b) room temperature of two P-AP states with opposite chirality configurations for same SAF sample, which become distinguishable due to presence of pinning site located within one core radius of particle center. Pinning effect survives at high temperature since core potential continues to have single minimum. (c) Corresponding micromagnetic simulation with pinning site 8~nm from particle center; two colors show EA-magnetization of two antiparallel chirality configurations.}
\label{nondeg}
\end{figure}

\begin{figure}[!t]
\includegraphics[width=3.1in]{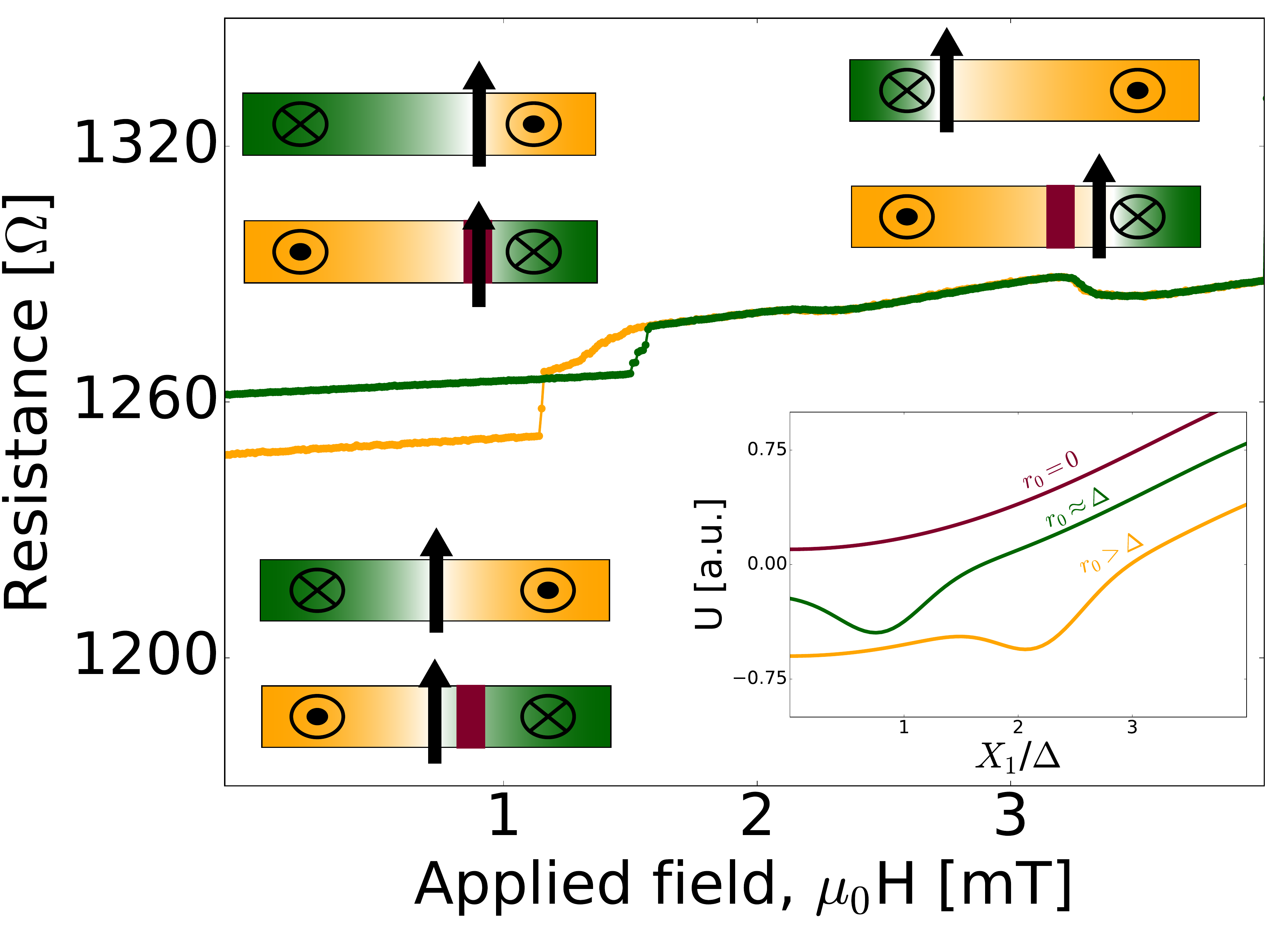}\\
\caption{Measured low temperature core hysteresis of P-AP vortex-pair state showing zero-field hysteresis due to pinning; green points represent decreasing field, yellow -- increasing. Schematics illustrate the core position at different points during sweep, with vortex cores coupling and decoupling while moving in and out of pinning site (indicated in red). Pinning site is located such as to yield second minimum in core potential. Pinning effect does not survive to high temperature due to thermally agitated transitions between two potential minima. Inset shows calculated core potential without pinning (red) and with pinning site located at two characteristic distances from particle center (green/yellow). Curves in inset are vertically offset for clarity.}
\label{zfh}
\end{figure}

Defects in magnetic films can pin a vortex core as it moves through the magnetic lattice of the material. Pinning of the vortex core can occur on intrinsic defects in the film, such as grain boundaries in polycrystalline films with varying local saturation magnetization, exchange stiffness, or magnetoelastic energy (strain).\cite{OHandley1999} Another possible source of pinning is the surface/interface roughness, leading geometrically to local variations in the vortex-core length.\cite{Chen2012a,Holmgren2018} The defects mainly interact with the sharp magnetic singularity of the vortex core in a vortex, and pin most effectively if their size is comparable to the core size.\cite{Chen2012a,Burgess2013} 

In a P-AP vortex pair, due to the strong core-core attraction in the coupled state, pinning of one core effectively pins both vortex cores (the pair). The specific effects of pinning differ depending on the location and energy of the pinning site. Three general cases can be distinguished. 

First, if the pinning site is roughly within one core radius ($\Delta \sim$ 10 nm in Py) of the geometrical center of the SAF particle and of similar strength to the core-core coupling, the minimum of the core potential at zero applied field is generally shifted from the center of the particle to reflect the presence of the core-pinning site. The potential does not split into two minima -- the coupled pair equilibrium position is shifted, to adjust to the new location of the potetial minimum. This results in a finite offset of the core pair towards the pinning site at zero field. The displacment of the pair is in the same direction regardless of the specific chirality state, so the change in the in-plane magnetization is of different sign for different chiralities. As a result, pinning can lift the chirality degeneracy and differentiate multiple P-AP states within the particle. A junction observed to have several P-AP states, shown in Fig.~\ref{nondeg}(a,b), is compared with a micromagetic simulation for two AP-chirality states, shown in Fig.~\ref{nondeg}(c). Since the minimum of the core potential is shifted slightly, without creating a second minimum, the core offset is insensitive to thermal agitation. Consequently, the splitting of the two chirality states observed at 77 K survives to room temperature (Fig.~\ref{nondeg}(a) and Fig.~\ref{nondeg}(b), respectively), even though the core-core hysteresis is thermally suppressed at RT.

The second characteristic case is where the pinning site is slightly more than one core-radius away from the center of the particle. The core potential now splits to create a second minimum, a distance away from the center of the particle. This results in hysteresis at zero field as the coupled pair can collectively be pinned or depinned, as shown in Fig.~\ref{zfh}. The inset to the figure illustrates the core potential, (\ref{Utot}), with an additional pinning energy
\begin{eqnarray}
U_{pin}=\beta e^{-\frac{(\mathbf{X}_1-\mathbf{X}_0)^2}{\Delta^2}}.
\label{Upin}
\end{eqnarray}
Here factor $\beta$ scales the pinning strength and $\mathbf{X}_0$ is the location of the pinning site in the bottom layer. Due to the highly-localized shape of the vortex core and a similarly localized and sized pinning site, their interaction is well modeled as having an exponentially decaying with distance pinning strength. Thermal fluctuations can smear the boundary between the two  stable states by agitating interstate core-pair transitions, so the pinning effect may not survive to higher temperatures.

Finally, the pinning site can be much further than one core size from the center of the particle, where it has no effect on the coupled core-core state. Instead the cores interact with the pinning site only in the decoupled state, where they behave effectively like isolated vortex cores. The effects of pinning in this case of an individual vortex core have been studied extensively in other publications.\cite{Burgess2013,Burgess2014,Chen2012a,Chen2012}


\section{\label{sec:con}Conclusions}
We have investigated the magnetostatic properties of a strongly coupled vortex pair with parallel cores and antiparallel chiralities as a function of various types of magnetic asymmetry often present in a multilayered nanopillar. Our analysis shows that thickness imbalance or dc-field biasing does not allow to differentiate via the standard resistive read-out the nominally degenerate	 vortex-pair states. Core pinning, on the other hand, is shown to differentiate in read-out degenerate chirality states, which at the same time can be insensitive to thermal fluctuations. These results should be useful for designing non-volatile vortex-based memory, where the information is stored in the topologically protected chirality states of the vortex pair. 
 
\section{\label{sec:ack}Acknowledgement}
Support from the Swedish Research Council (VR Grant No. 2014-4548, 2018-03526) is gratefully acknowledged.

\bibliographystyle{unsrt}
\bibliography{refs}

\end{document}